\documentclass[aps,twocolumn,showpacs,superscriptaddress,preprintnumbers,floatfix,amsmath,amssymb]{revtex4-1}

\usepackage{amssymb}
\usepackage{graphicx}
\usepackage{dcolumn}
\usepackage{bm}
\usepackage{color}
\hfuzz=\maxdimen
\begin{document}

\title{Superconductivity in quasi-one-dimensional K$_2$Cr$_3$As$_3$ with significant electron correlations}

\author{Jin-Ke Bao}
\affiliation{Department of Physics, Zhejiang University, Hangzhou
310027, China}
\author{Ji-Yong Liu}
\affiliation{Department of Chemistry, Zhejiang University, Hangzhou
310027, China}
\author{Cong-Wei Ma}
\affiliation{Department of Physics, Zhejiang University, Hangzhou
310027, China}
\author{Zhi-Hao Meng}
\affiliation{Department of Physics, Zhejiang University, Hangzhou
310027, China}
\author{Zhang-Tu Tang}
\affiliation{Department of Physics, Zhejiang University, Hangzhou
310027, China}
\author{Yun-Lei Sun}
\affiliation{Department of Physics, Zhejiang University, Hangzhou
310027, China}
\author{Hui-Fei Zhai}
\affiliation{Department of Physics, Zhejiang University, Hangzhou
310027, China}
\author{Hao Jiang}
\affiliation{Department of Physics, Zhejiang University, Hangzhou
310027, China}
\author{Hua Bai}
\affiliation{Department of Physics, Zhejiang University, Hangzhou
310027, China}
\author{Chun-Mu Feng}
\affiliation{Department of Physics, Zhejiang University, Hangzhou
310027, China}

\author{Zhu-An Xu}
\affiliation{Department of Physics, Zhejiang University, Hangzhou
310027, China} \affiliation{State Key Lab of Silicon Materials,
Zhejiang University, Hangzhou 310027, China} \affiliation{Collaborative Innovation Centre of Advanced Microstructures, Nanjing 210093, China}

\author{Guang-Han Cao} \email[Correspondence should be sent to: ]{ghcao@zju.edu.cn}
\affiliation{Department of Physics, Zhejiang University, Hangzhou
310027, China} \affiliation{State Key Lab of Silicon Materials,
Zhejiang University, Hangzhou 310027, China} \affiliation{Collaborative Innovation Centre of Advanced Microstructures, Nanjing 210093, China}

\date{\today}

\begin{abstract}
We report the discovery of bulk superconductivity (SC) at 6.1 K in a quasi-one-dimensional (Q1D) chromium pnictide K$_2$Cr$_3$As$_3$ which contains [(Cr$_3$As$_3$)$^{2-}$]$_{\infty}$ double-walled subnano-tubes with face-sharing Cr$_{6/2}$ (As$_{6/2}$) octahedron linear chains in the inner (outer) wall. The material has a large electronic specific-heat coefficient of 70$\sim$75 mJ K$^{-2}$ mol$^{-1}$, indicating significantly strong electron correlations. Signature of non-Fermi liquid behavior is shown by the linear temperature dependence of resistivity in a broad temperature range from 7 to 300 K. Unconventional SC is preliminarily manifested by the estimated upper critical field exceeding the Pauli limit by a factor of three to four. The title compound represents a rare example that possibly unconventional SC emerges in a Q1D system with strong electron correlations.\\

Subject Areas: Condensed Matter Physics, Superconductivity,
Strongly Correlated Materials
\end{abstract}

\maketitle

\section{\label{sec:level1}Introduction}
The electron correlations and the reduced dimensionality, as central issues in contemporary condensed matter physics, play an important role in producing novel superconductivity (SC) in crystalline materials. This is prominently exemplified in the quasi-two-dimensional (Q2D) cuprates\cite{bednorz}, strontium ruthenate\cite{maeno} and ferroarsenides\cite{hosono}, in which the correlated $d$-electrons are believed to be essential for the appearance of unconventional SC. As the dimensionality is further reduced to the Q1D scenario, however, SC occurs infrequently mainly because of Peierls instability\cite{peierls,smaalen}.  Representative examples of Q1D superconductors include the earlier organic Bechgaard salts\cite{jerome} and the purple molybdenum bronze Li$_{0.9}$Mo$_6$O$_{17}$\cite{greenblatt}, where the C-2$p$ electrons and the Mo-4$d$ electrons are responsible for SC. Since 3$d$-transition elements generally bear stronger electron correlations, it is of great interest if a 3$d$-element-based Q1D compound would superconduct as well. The Cu-based ¡°ladder¡± material, as a crossover from 1D to 2D, was theoretically predicted\cite{dagotto} and, experimentally confirmed\cite{uehara} to superconduct, although under high pressures. Unfortunately, the SC actually locates in the Q2D regime due to a dimension crossover\cite{nagata}.

It was revealed that both cuprate and iron-based high-temperature superconductors may bear 1D feature such as stripes\cite{tranquada} and nematicity\cite{chu}, albeit of their Q2D crystal structures. Thus, investigations on SC in Q1D correlated electron systems may shed light on the long-standing puzzle of unconventional superconducting mechanism because of the inherent simplicity of one dimensionality. Additional interest for pursuing Q1D material with interacting electrons comes from the possible realization of Luttinger liquid in which an exotic spin-charge separation is expected\cite{voit}.

Here we report synthesis and characterizations of a chromium arsenide, K$_2$Cr$_3$As$_3$, which explicitly shows a Q1D crystal structure featured by well separated [(Cr$_3$As$_3$)$^{2-}$]$_{\infty}$ chains in the crystalline lattice. Strong electron correlations are indicated by the large electronic specific-heat coefficient. Remarkably, the new material hosts novel normal-state and superconducting properties which point to unconventional SC.

\section{\label{sec:level2}Experimental Methods}

\begin{figure*}
\includegraphics[width=14cm]{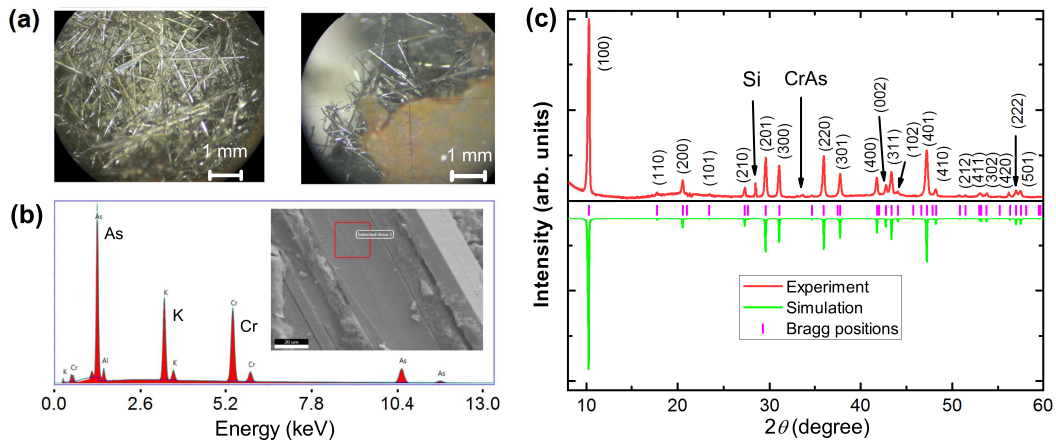}
\caption{\label{fig1} Characterizations of the K$_2$Cr$_3$As$_3$ samples. (a) Morphology of a batch of the as-grown crystals under an optical microscope (left: top view; right: side view). (b) A typical energy dispersive X-ray spectrum with electron beams focused on the selected area (marked in the inset) of the as-grown crystals. Small amount of the element Al comes from the sample holder. (c) Powder X-ray diffraction of the polycrystalline samples indexed by the hexagonal unit cell determined from the single-crystal diffractions. Si was added as an internal standard. Only small amount of the CrAs impurity was identified. The simulated pattern (with the consideration of preferential orientations of crystalline grains) well reproduces the experimental data.}
\end{figure*}

\paragraph{Sample's synthesis} Single crystals of K$_2$Cr$_3$As$_3$ were grown by spontaneous nucleation in the high-temperature solution with a self flux of KAs under vacuum. First, KAs (CrAs) was prepared by the reaction of the  stoichiometric mixture of potassium pieces (99.95\%), chromium powder (99.95\%) and arsenic powder (99.999\%) at 473 (973) K for 16 h in evacuated quartz tubes. Cautions should be taken to avoid the possible violent chemical reactions, and it is highly recommended that the initial heating should be very slow. Second, the presynthesized CrAs and KAs powders were mixed in a molar ratio of 1:6, namely, K:Cr:As =6:1:7, and the mixtures were loaded in an alumina crucible. The alumina crucible (covered with a cap) was jacketed in a Ta tube, and the welded Ta tube was finally sealed in an evacuated quartz ampoule. The sample-loaded quartz ampoule was heated up to 1273 K holding for 24 h in a furnace, followed by cooling down to 923 K at a rate of 2 K/h. Shiny needle-like crystals with a typical size of 2$\times$0.1$\times$0.05 mm were harvested, as photographed in Fig.~\ref{fig1}(a). The polycrystalline sample was synthesized by a solid state reaction in vacuum using the three elements as the starting materials. The stoichiometric ratio of K$_2$Cr$_3$As$_3$ was taken. After the initial reaction, the mixtures were ground, and then pressed into a pellet. The pellet was finally sintered in vacuum at 973 to 1173 K, holding for 24 h. The resulted sample was dense, and it was suitable for the electrical resistivity and specific heat measurements. Note that most of the procedures were carried out in an argon-filled glovebox with the water and oxygen content below 0.1 ppm. The synthesized samples are very reactive in air, and easily deteriorate at ambient condition. Thus, it is important to avoid exposure to air as far as possible while handling samples.

\paragraph{Composition determination} The chemical composition of the as-grown single crystals was measured by energy-dispersive X-ray spectroscopy (EDS) with an AMETEK$\copyright$ EDAX (Model Octane Plus) spectrometer, equipped in a field-emission scanning electron microscope (Hitachi S-4800). The typical EDS spectrum was shown in Fig.~\ref{fig1}(b). Through multiple measurements with a control experiment [see details in the Supplemental Material (SM)\cite{SM}], the chemical composition of the as-grown crystals were determined to be K$_{1.82(19)}$Cr$_{3.00}$As$_{2.99(12)}$, which is consistent with the ideal formula of K$_2$Cr$_3$As$_3$ derived from the crystal structure.

\paragraph{Structural determination} The crystal structure was determined by single-crystal X-ray diffractions on an Xcalibur, Atlas, Gemini ultra diffractometer. Details of the experiment and analysis are presented in SM~\cite{SM}. The result indicates that the most probable structure at room temperature and below has a space group of $P\overline{6}m2$ (No. 187). The obtained crystallographic structural data as well as the related bond lengths and bond angles are tabulated in Tables S3 to S5\cite{SM}.

Powder X-ray diffraction was carried out at room temperature on a PANalytical x-ray diffractometer (Empyrean Series 2) with a monochromatic CuK$_{\alpha1}$ radiation. To avoid the reaction with air, a thin layer of Apiezon $N$-grease was carefully coated on the sample's surface. Small amount of Si powder was added as the internal standard reference material. As shown in Fig. 1(c), the powder X-ray diffraction peaks, both in positions and densities, are quantitatively consistent with the simulated profile using the crystal structure data in Tables S3 and S4. This clearly indicates the same phase for the single crystals and polycrystals.

\paragraph{Physical property measurements} We employed a Magnetic Property Measurement System (MPMS-5, Quantum Design) to measure the magnetic property. Both single crystals (a bundle of the crystal sticks) and bulk polycrystals were measured under an applied field of $H$ = 10 Oe with zero-field-cooling and field-cooling protocols. The electrical resistivity and the specific-heat capacity were measured for the polycrystalline samples, using standard four-terminal method and relaxation technique, respectively, on a Physical Properties Measurement System (PPMS-9, Quantum Design). The heat capacity from the sample holder and grease was deducted. Note that our preliminary attempt to measure on a stick of single crystal was unsuccessful and/or unsatisfactory because of its tiny mass (for the magnetic and specific heat measurements) or its high reactivity (for the resistivity measurement).

\section{\label{sec:level3}Results and discussions}

\subsection{\label{subsec:level1}Crystal structure}

\begin{figure}
\includegraphics[width=8cm]{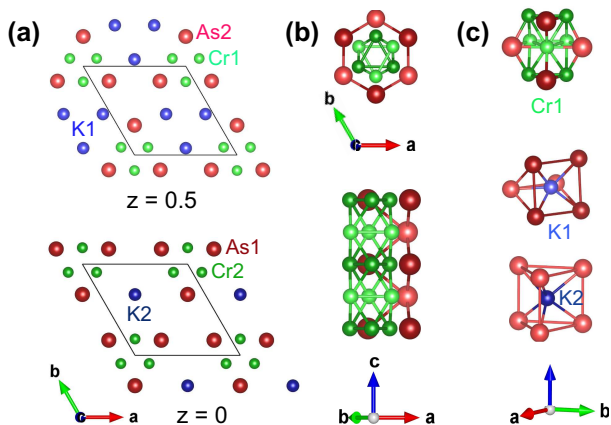}
\caption{\label{fig2} Crystal structure of K$_2$Cr$_3$As$_3$. (a) The constituted atoms locate in the two crystalline planes with $z$ = 0 and $z$ = 0.5. (B) The structure of [(Cr$_3$As$_3$)$^{2-}$]$_{\infty}$ double-walled subnano-tubes (top: top view; bottom: side view with removal of the outer wall in the left front). (C) Chemical bonding of Cr1 (top), K1 (middle) and K2 (bottom).}
\end{figure}

K$_2$Cr$_3$As$_3$ crystallizes in a hexagonal lattice with $a$ = 9.9832(9) \AA, $c$ = 4.2304(4) \AA, and the most probable space group of $P\overline{6}m2$ (No. 187) at room temperature (13). The constituent atoms locate in the two crystalline planes with $z$ = 0 and $z$ = 0.5, as illustrated in Fig. ~\ref{fig2}(a). The prominent structural unit is the 1D negatively charged 'chains', shown in Fig.~\ref{fig2}(b). The [(Cr$_3$As$_3$)$^{2-}$]$_{\infty}$ chains are actually double-walled subnano-tubes (DWSTs) of 0.58 nm in outer diameter, in which face-sharing Cr$_6$ (As$_6$) octahedron tubes constitute the inner (outer) wall. The Cr$_{6/2}$ 'monomer' is basically a regular octahedron with Cr$-$Cr bond lengths varying from 2.61 to 2.69(1) \AA. These Cr$-$Cr bond distances are close to that (2.50 \AA) in Cr metal, but much larger than twice (1.46 \AA) of the radius of Cr$^{2+}$, indicating virtually metallic bonding among Cr atoms. Apart from the metallic bonding with six neighbouring Cr atoms, each Cr atom also behaves like a cation (the apparent valence of Cr is +2.33 if assuming K$^+$ and As$^{3-}$, which ionically bonds with four arsenic anions [see the top of Fig.~\ref{fig2}(c)]. The DWSTs are well separated by columns of the K counterions, hence the inter-DWST coupling is expected to be much weaker than the intra-DWST interactions. There are two distinct crystallographic sites for potassium. K1 is coordinated with As$_6$ triangular prisms horizontally [the middle of Fig.~\ref{fig2}(c)]. K2 is also bonded with As$_6$ triangular prisms, but vertically [see the bottom of Fig. 2(c)] and, there are six additional arsenic anions surrounded farther within the same $ab$ plane (not shown).

After solving the crystal structure, we became aware of a similar structure in the $Ak$Mo$_3$$Ch_3$ or $Ak_2$Mo$_6$$Ch_6$ ($Ak$ = alkli metals, Tl, and In; $Ch$ = chalcogen elements) series\cite{potel,honle}. Although the 1D structural unit [(Mo$_3$$Ch_3$)$^{-}$]$_{\infty}$ is isostructural to the DWST of [(Cr$_3$As$_3$)$^{2-}$]$_{\infty}$, the former rotates a small angle along the $c$ axis, relative to the orientation of the latter. This leads to a different space group, $P6_{3}m$ (No. 176), for $Ak$Mo$_3$$Ch_3$. The $Ak$ atoms only occupy the (1/3, 2/3, 0) and (2/3, 1/3, 1/2) sites, which gives 1:3:3 stoichiometry. Another important difference lies in the slight distortion of the [(Cr$_3$As$_3$)$^{2-}$]$_{\infty}$ DWSTs, accompanied with the absence of the inversion center in K$_2$Cr$_3$As$_3$.

\subsection{\label{subsec:level2}Superconductivity}

\begin{figure*}
\includegraphics[width=14cm]{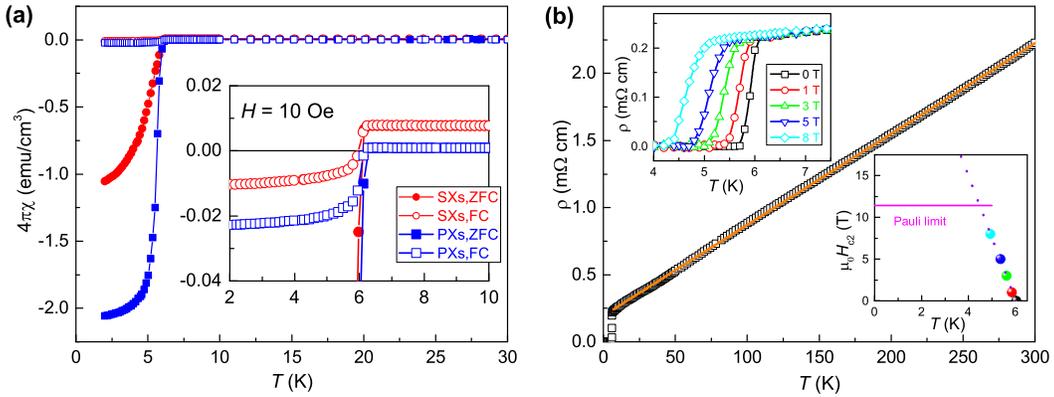}
\caption{\label{fig3} Superconductivity in K$_2$Cr$_3$As$_3$. (a) Temperature dependence of dc magnetic susceptibility for K$_2$Cr$_3$As$_3$ single crystals (SXs) and polycrystals (PXs). The demagnetization effect was not taken into consideration. ZFC and FC denote zero-field cooling and field cooling, respectively. The inset zooms in the FC curves. (b) Temperature dependence of resistivity for the K$_2$Cr$_3$As$_3$ polycrystalline sample. The orange line is the linear fit for the normal state. The upper left inset shows the superconducting transitions under different magnetic fields up to 8 T, from which the upper critical field ($H_{c2}$) was derived (shown in the lower right inset). The Pauli limit for $H_{c2}$ is marked by the horizontal line.}
\end{figure*}

Figure ~\ref{fig3}(a) shows temperature dependence of magnetic susceptibility, $\chi(T)$, for a bundle of the crystal ¡®sticks¡¯ (the sticks were mostly parallel to the external field) and the bulk polycrystalline sample of K$_2$Cr$_3$As$_3$. A sharp diamagnetic transition can be seen at $T_{\text{c}}$ = 6.1 K. The volume fractions (scaled by 4$\pi\chi$) of magnetic shielding, measured in the zero-field-cooling mode, exceed 100\% at 2 K (they are approximately to be 100\% after a correction of the demagnetization effect), suggesting bulk SC. On the other hand, the volume fractions of magnetic repulsion, reflected by the field-cooling data, are much smaller mainly because of the magnetic flux pinning in the process of cooling down under magnetic fields. The sharpness of the superconducting transitions, which is also shown in the following measurements, suggests sufficiently strong inter-chain couplings albeit of the obvious Q1D structure.

Figure ~\ref{fig3}(b) shows temperature dependence of electrical resistivity, $\rho(T)$, measured with the polycrystalline sample\cite{polycrystals}. Remarkably, the resistivity follows with $\rho(T) = \rho_{0} + AT$ with $\rho_{0}$ = 0.195 m$\Omega$ cm and $A$ = 0.0067 m$\Omega$ cm K$^{-1}$ from 7 to 300 K\cite{linear}. This is reminiscent of the $T$-linear resistivity in optimally-doped cuprate superconductors\cite{gurvitch}, iron pnictide superconductors\cite{jiang}, and $f$-electron systems\cite{stewart}, which is regarded as one of the hallmarks of non-Fermi liquid behavior\cite{khalifah}. Owing to the Q1D characteristic here, the origin of the $T$-linear resistivity could be related to the possible Luttinger liquid\cite{voit,anderson}. SC is confirmed by the resistivity drop below 6.1 K, as clearly seen in the top-left inset of Fig. ~\ref{fig3}(b).

With applying magnetic field, the superconducting transition moves slowly to lower temperatures. The upper critical field ($H_{c2}$) was then determined using the common criteria of 90\% of the extrapolated normal-state resistivity at the superconducting transition. The lower right inset in Fig.~\ref{fig3}(b) plots the obtained $H_{c2}$ as a function of temperature, showing that $H_{c2}$ increases very steeply with decreasing temperature. The absolute value of the initial slope, $-\mu_{0}$(d$H_{c2}$/d$T$)$_{T\rightarrow T_{c}}$, is as high as 7.43 T/K. By a simple linear extrapolation, the zero-temperature upper critical field, $\mu_{0}H_{c2}(0)$, is estimated to be 44.7 T. If applying the Werthammer-Helfand-Hohenberg model\cite{WHH}, in which only orbital effect is taken into account, the $\mu_{0}H_{c2}(0)$ value is then estimated to be about 32 T. Here we note that the $H_{c2}$ value would be even larger if the magnetic field is along the $c$ axis of a crystal. On the other hand, the Pauli paramagnetic limit for the upper critical field is $\mu_{0}H_{\text{\text{P}}}$ = 1.84 $T_{\text{c}}$ $\approx$ 11 T in the case of isotropic full superconducting gap without spin-orbit coupling)\cite{clogston,chandrasekhar}. Therefore, the $H_{c2}(0)$ value could be over 300-400\% of the Pauli limit, which implies a novel spin-triplet Cooper pairing like that in the $p$-wave SC in Sr$_2$RuO$_4$\cite{nelson}. Further investigations using other techniques such as nuclear magnetic resonance are expected to be helpful to clarify this interesting issue.

\begin{figure}
\includegraphics[width=8cm]{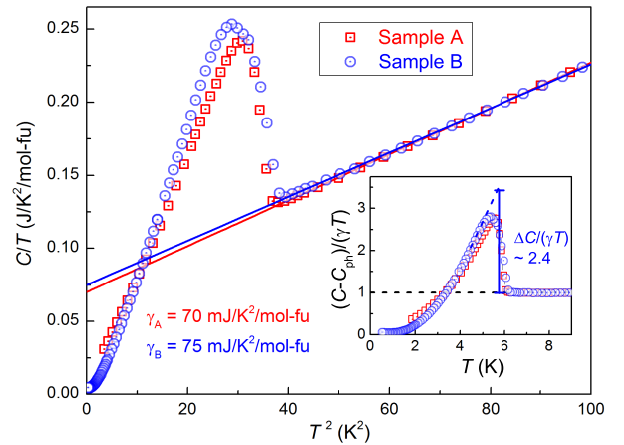}
\caption{\label{fig4} Low-temperature specific heat for the K$_2$Cr$_3$As$_3$ polycrystals. The main figure plots $C/T$ vs. $T^2$, which extracts the contributions from phonons ($C_\text{ph}$ = $\beta T^3$) and electrons ($C_\text{e}$ = $\gamma T$), and yields electronic specific-heat coefficients of $\gamma$ = 70.2 and 75.0 mJ K$^{-2}$ mol-fu$^{-1}$, respectively, for Sample A and Sample B. The inset plots ($C-\beta T^3$)/($\gamma T$) as a function of temperature.}
\end{figure}

Temperature dependence of specific heat, $C(T)$, may supply important information for a superconductor. Fig.~\ref{fig4} shows low-$T$ specific-heat data for the K$_2$Cr$_3$As$_3$ polycrystals. The $C/T$ vs $T^2$ plot allows us to extract the contributions from electrons ($C_\text{e}$ = $\gamma T$) and phonons ($C_\text{ph}$ = $\beta T^3$). The linear fit for the data from 6.4 to 10 K yields an electronic specific-heat coefficient $\gamma$ = 70.2 (for Sample A) or 75.0 (for Sample B) mJ K$^{-2}$ mol-fu$^{-1}$  (fu refers to formula unit), equivalent to 23.4 or 25.0 mJ K$^{-2}$ mol-Cr$^{-1}$. This $\gamma$ value exceeds twice of that (9.1 mJ K$^{-2}$ mol-Cr$^{-1}$) of the related three-dimensional correlated metal CrAs\cite{wu}, indicating enhanced electron correlations in the Q1D K$_2$Cr$_3$As$_3$. The experimental density of states at Fermi level, $N(E_\mathrm{F}) = 3\gamma/(\pi k_\mathrm{B})^2 = 30\sim32$ eV$^{-1}$ fu$^{-1}$ ($k_\mathrm{B}$ is Boltzmann constant), is over three times of the ¡®bare¡¯ density of states (8.58 eV$^{-1}$ fu$^{-1}$) from the first-principles calculations\cite{jh}, which beyond the explanation by electron-phonon interactions and, quantum fluctuations might be involved. With the fitted $\beta$ value of 1.57 (Sample A) or 1.51 (Sample B) mJ K$^{-4}$ mol$^{-1}$, and using the formula $\theta_{\text{D}}=[(12/5)NR\pi^{4}/\beta]^{1/3}$, the Debye temperature can be calculated to be 215 and 218 K.

Below 6 K, a characteristic specific-heat jump ($\Delta C$) due to the superconducting transition shows up, further confirming the bulk SC. Actually, $\Delta C$ comes from the change in the electronic part, $C_\text{e} \approx C - \beta T^3$ and, for revealing the $C_\text{e}(T)$ behavior in the superconducting state, $(C - \beta T^3)$ is normalized by the normal-state specific heat $\gamma T$ (the inset of Fig.~\ref{fig4}). As is seen, the dimensionless specific-heat jump at $T_{\text{c}}$, [$\Delta C/(\gamma T_{\text{c}})$], is as high as 2.4 for Sample B, significantly larger than the theoretical value (1.43) of the well-known BCS theory, suggesting a strong coupling scenario. By the entropy conserving construction, the thermodynamic transition temperature is determined to be 5.75 K, coincident with the zero-resistance temperature. In general, the $C_\text{e}(T)$ dependence may give the information of low-energy quasiparticles that reflects the superconducting pairing symmetry. However, we found that the ($C-\beta T^3$)/($\gamma T$) data show an upturn below 1 K, which is probably due to a Schottky-like anomaly from the related nuclei and/or impurities. This extra contribution prevent us from fitting the data to give a reliable conclusion.

\section{\label{sec:level4}Concluding Remarks}

We have discovered superconductivity at 6.1 K under ambient pressure in a new Cr-based material K$_2$Cr$_3$As$_3$. The crystal structure is characterized by the [(Cr$_3$As$_3$)$^{2-}$]$_{\infty}$ "chains", well separated by K$^{+}$ cations, which makes title compound as an explicitly Q1D material. Although the $T_{\text{c}}$ is not impressively high, to the best of our knowledge, however, it is among the highest in Q1D systems. The material also possesses a large electronic specific-heat coefficient of 70$\sim$75 mJ K$^{-2}$ mol$^{-1}$, indicating significantly strong electron correlations.

As a rare example of Q1D superconductor with significant electron correlations, K$_2$Cr$_3$As$_3$ also shows peculiar physical properties. The $T$-linear resistivity in a broad temperature range suggests a non-Fermi-liquid normal state, which could be related to a quantum criticality and/or realization of Luttinger liquid. The estimated $\mu_{0}H_{c2}(0)$ value, even for the polycrystalline sample, is three to four times of the Pauli limit (in comparison, the $\mu_{0}H_{c2}(0)$ values of the Q1D 4.2 K superconductor TlMo$_3$Se$_3$ are 5.8 T and 0.47 T, respectively, for field parallel and perpendicular to the $c$ axis\cite{petrovic}). This is obviously beyond the explanation of spin-orbit scattering and multiband effect, and thus a spin-triplet pairing is likely.

The appearance of SC in Q1D K$_2$Cr$_3$As$_3$ is an unexpected result. First, the tendency of Peierls transition in Q1D metals always prevents it from SC. For example, most members of the aforementioned Q1D compounds $Ak$Mo$_3$$Ch_3$\cite{potel,honle}, including that of $Ak$ = K, do not show SC (exceptions are $Ak$ = Tl and In). Second, chromium is the only element metal that shows antiferromagnetism above room temperature, and it is one of the minor metallic elements that does not superconduct even under high pressures. While a few alloys containing Cr superconduct\cite{matthias}, SC in Cr-based compounds was not reported until very recent observation of SC at $\sim$2 K in a three-dimensional compound CrAs under high pressures\cite{LJL,kotegawa}. Therefore, K$_2$Cr$_3$As$_3$ is the first Cr-based superconductor at ambient pressure. The occurrence of SC in this Q1D Cr-based material calls for further investigations.
\\

\begin{acknowledgments}
We would like to thank H. Chen for his constructive suggestions on the single-crystal diffraction experiment. Thanks are also due to F. C. Zhang, C. Cao, Y. Zhou, J. H. Dai, H. Q. Yuan for helpful discussions. This work was supported by the Natural Science Foundation of China (No. 11190023), the National Basic Research Program (No. 2011CBA00103), and the Fundamental Research Funds for the Central Universities of China.
\end{acknowledgments}


\begin{thebibliography}{33}
\bibitem [{pol()}]{polycrystals}%
  \BibitemOpen
  \href@noop {} {}\bibinfo {note} {Our preliminary measurements on a single
  stick of crystal were unreproducible, accompanied with deterioration
  of superconductivity. It was probably owing to the use of silver paste (for
  making electrodes) which could react with the whole tiny crystal. However, for the polycrystalline pellets, the measurements were able to be reproducible, primarily because only a surface layer would be damaged by the silver paste.}\BibitemShut {Stop}%
\bibitem [{lin()}]{linear}%
  \BibitemOpen
  \href@noop {} {}\bibinfo {note} {We found that exposure of sample in
  air for a short time could lead to deviation of the linearity, accompanied with a reduction of $T_\mathrm{c}$. This leads us believe that the linear-$T$ resistivity is an intrinsic property of the K$_2$Cr$_3$As$_3$ polycrystals, which mainly reflects the $\rho_{\|}(T)$ (resistivity along the chain direction) behavior. Nevertheless, the linear $\rho(T)$ might also be due to an incidental "polycrystal effect", i.e., the as-measured linear $\rho(T)$ is a combined result of $\rho_{\|}(T)$, $\rho_{\bot}(T)$, and the grain boundaries. Thus, the intrinsic $\rho_{\|}(T)$ and $\rho_{\bot}(T)$ need to be examined by using the single-crystal sample without any deterioration.}\BibitemShut {Stop}%
\end{thebibliography}
\end{document}